\documentstyle[pre,aps,twocolumn,psfig]{revtex}     % journal page size

\newcommand{\figutwo}[4]{
                         \nobreak \hbox to \textwidth { \leavevmode 
                         \epsfxsize=#1 \epsffile{#2}
                         \hss \leavevmode 
                         \epsfxsize=#3 \epsffile{#4} }
                         \nobreak
                        }
\newcommand{\figu}[2]{\nobreak \hbox to \textwidth {
                      \centerline{\leavevmode
                      \epsfxsize=#1 \epsffile{#2}
                      }} \nobreak 
                     }

\newcommand{\bge}{\begin{equation}}
\newcommand{\ede}{\end{equation}}
\newcommand{\ba}{\begin{array}}
\newcommand{\ea}{\end{array}}

\newcommand{\f}{\frac}

\newcommand{\no}{\nonumber}
\input psfig.sty

\newcommand{\rf}     [1] {~\cite{#1}}

\newcommand{\reffig} [1] {fig.~\ref{#1}}

\newcommand{\beq}{\begin{equation}}
\newcommand{\continue}{\nonumber \\ }

\newcommand{\eeq}{\end{equation}}
\newcommand{\ee}[1] {\label{#1} \end{equation}}
\newcommand{\bea}{\begin{eqnarray}}

\newcommand{\eea}{\end{eqnarray}}
\newcommand{\barr}{\begin{array}}
\newcommand{\earr}{\end{array}}

\newcommand{\evOper}{evolution oper\-ator}

\newcommand{\fd}{spec\-tral det\-er\-min\-ant}

\newcommand{\Lop}{{\cal L}}	   % evolution operator
\begin{document}
\draft{
\title{Asymptotics of high order noise corrections}
\author{Niels S\o ndergaard}
\address{
Department of Physics \&\ Astronomy, Northwestern University \\
2145 Sheridan Road, Evanston, Illinois 60208}
\author{Gergely Palla and G\'abor Vattay}
\address{Department of Physics of Complex Systems, E\"otv\"os University\\
P\'azm\'any P\'eter s\'etany 1/A,
H-1117 Budapest, Hungary   } 
\author{ Andr\'e Voros} 
\address{
CEA, Service de Physique Th\'eorique de Saclay\\
F-91191 Gif-sur-Yvette CEDEX (France)}

\date{\today}

\maketitle

\begin{abstract}
We consider an evolution operator for a discrete Langevin equation with a strongly hyperbolic 
classical dynamics and noise with finite moments. Using a perturbative expansion of the evolution operator we calculate high order corrections to its trace in the case of a quartic map and Gaussian noise. The leading contributions  come from the period one orbits of the map. The asymptotic behaviour is investigated and is found to be independent up to a multiplicative constant of the distribution of noise.

\end{abstract}

\section{Introduction}

In natural phenomena stochastic processes of various strength  always have an influence.
In the three preceding papers\rf{noisy_Fred,conjug_Fred,diag_Fred}
 the effects of noise on measurable
properties such as dynamical averages\cite{bene} in classical chaotic dynamical
systems  were systematically accounted.
In order to construct a statistical theory of dynamical systems one introduces 
densities of particles governed by a corresponding evolution operator.
For a repeller the leading eigenvalue of this operator $\Lop$ 
yields a physically measurable property of the dynamical system,
the escape rate from the repeller.
In the case of deterministic flows, the periodic orbit theory
yields explicit and numerically efficient formulas for the spectrum of $\Lop$ as zeros
of its {\fd}\rf{QCcourse}.

The theory developed is  closely related to the semiclassical expansions based
on Gutzwiller's formula for the trace in terms of classical periodic
orbits\rf{gutbook} in that both are perturbative theories (in the noise
strength or $\hbar$) derived from saddle point expansions of a path
integral containing a dense set of unstable stationary points (typically
periodic orbits).  The analogy with quantum mechanics and
field theory is made explicit in\rf{noisy_Fred} where
Feynman diagrams are used to find the lowest nontrivial noise corrections
 to the escape rate.

An elegant method, inspired by the classical perturbation theory of
celestial mechanics, is that of smooth conjugations\rf{conjug_Fred}.
In this approach
the neighbourhood of each saddle point is flattened by an appropriate
coordinate transformation, so the focus shifts from the
 original dynamics
to the properties of the transformations involved. 
The expressions obtained for perturbative corrections in this approach
are much simpler than those found
from the equivalent Feynman diagrams.  Using these techniques, we were
able to extend the stochastic perturbation theory to the fourth order
 in the
noise strength.

In \rf{diag_Fred}  we develop
a third approach, based on construction of an explicit matrix
representation of the stochastic evolution operator. 
The numerical implementation requires a
truncation to finite dimensional matrices,
and is less elegant than the smooth conjugation method,
but makes it possible to reach  up to order eight in expansion orders.
As with the previous formulations, it retains the periodic orbit structure,
thus inheriting valuable information about the dynamics.

The corrections to the escape rate were found to be a divergent series in the 
noise expansion parameter. This reflects  that the corrections are calculated using the
so-called cumulant expansion from other divergent quantities, the traces of the evolution operator $\Lop^n$ \rf{diag_Fred}.
In this article the focus is on  the high order 
corrections for the special case of the first trace,  tr$(\Lop)$. 
For the asymptotic study  we have to calculate a sufficient number of corrections and here we 
are led naturally to a contour integral method. Next the asymptotic behaviour is extracted by the method of steepest descent.

\section{The stochastic evolution operator}
In this section we introduce the noisy repeller and its evolution operator. 

An individual trajectory in presence of additive noise is generated
by iterating 
\beq
x_{n+1}=f(x_{n})+\sigma\xi_{n} 
\,,
\ee{mapf(x)-Diag}
where $f(x)$ is a map, 
$\xi_n$ a random variable with the normalized
distribution $p(\xi)$, 
and $\sigma$ parametrizes the noise strength.
In what follows we shall assume that the mapping $f(x)$ is
one-dimensional and expanding, and that the $\xi_n$ are uncorrelated.
A density of trajectories $\phi(x)$ evolves with time on the average as
\beq
\phi_{n+1}(y) =
\left(
\Lop
\circ
\phi_{n}\right)(y)
= \int dx \, \Lop(y,x) \phi_{n}(x)
\ee{DensEvol}
where the $\Lop$ {\evOper} has the general form of
\bea
\Lop(y,x) &=& \delta_\sigma(y-f(x)) ,\label{OpOverNoise}
	\continue
  \delta_\sigma(x)
  &=& \int \delta(x-\sigma \xi) p(\xi) d\xi 
  \,=\, \frac{1}{\sigma} p\left( \frac{x}{\sigma} \right)
\,.
\label{oper-Diag}
\eea
 For the calculations
in this paper Gaussian weak noise is assumed. In the perturbative limit,$\sigma \rightarrow 0$, the evolution operator becomes
\bea
\Lop(x,y)&=&\f{1}{\sqrt{2\pi}\sigma}e^{-\f{(y-f(x))^2}{2\sigma^2}}  \\
 &\sim &\sum_{n=0}^{\infty}a_{2n}\sigma^{2n}\delta^{(2n)}(y-f(x)),
\eea
where $\delta^{(2n)}$ denotes the $2n$-th derivative of the delta distribution and
$a_{2n}=(2^nn!)^{-1}$ or $a_{2n}= \f{m_{2 n}}{(2 n)!}$ for general noise with finite n'th moment $m_n$. Strictly speaking we here only consider distributions with vanishing odd moments, but the formulas below can easily be generalized. The perturbative form of the evolution operator is obtained formally from $(\ref{OpOverNoise})$ by Taylor expansion or 
by doing a saddle point integral discarding subdominant terms. In particular there may exist
stationary points which do not correspond to periodic orbits. However, these will give exponentially
small contributions and are therefore omitted in the weak noise limit $\sigma \rightarrow 0$. 
The map considered here is the same as in our previous papers, a quartic map on the
$(0,1)$ interval given by
\bea
f(x)=20\left[\f{1}{16}-\left(\f{1}{2}-x\right)^4\right].
\eea

\section{Residue method for the trace}
The trace of the  evolution operator in the weak-noise limit is obtained from
\bea
\mbox{Tr}{\cal L }=\int dx \sum_{n=0}^{\infty}{a_{2 n}\sigma^{2n}
\delta^{(2n)}(x-f(x))}. \label{trace}
\eea
By expressing the delta function as 
\bea
\delta(x)&=&\lim_{\epsilon\rightarrow 0}\f{1}{\pi}\mbox{Im}\f{1}{x-i\epsilon}=\no \\
& &\lim_{\epsilon\rightarrow 0}\f{1}{2\pi i}\left(
\f{1}{x-i\epsilon}-\f{1}{x+i\epsilon}\right),
\label{delta}
\eea
we reformulate expression ($\ref{trace}$) as
\bea
\mbox{Tr}{\cal L}&=&\lim_{\epsilon\rightarrow 0}\sum_{n=0}^{\infty}
\f{(2n)!a_n\sigma^{2n}}{2\pi i}\left[\int dx\f{1}{(x-f(x)-i\epsilon)^{2n+1}}\right.
\nonumber \\ & &\left.-\int dx\f{1}{(x-f(x)+i\epsilon)^{2n+1}}\right],
\label{reint}
\eea 
obtaining just the result as in \rf{Dingle} section 5.3 equation (19) p.119.

The integrals in ($\ref{reint}$) have poles close to the fixpoints of the map,
which are defined by 
\bea
x^*-f(x^*)=0.
\label{fixp}
\eea
For the map in question, a quartic, there are four points in the
complex plane which satisfy this equation: two of them are on
 the real axis: one at $x_0=0$, the other $x_1$ close to unity, as
 it can be seen on figure ($\ref{figmap}$). The two real fix points correspond
to orbits 0 and 1 in the language of symbolic dynamics \rf{QCcourse} .
\begin{figure}[hbt]
\centerline{\strut\psfig{figure=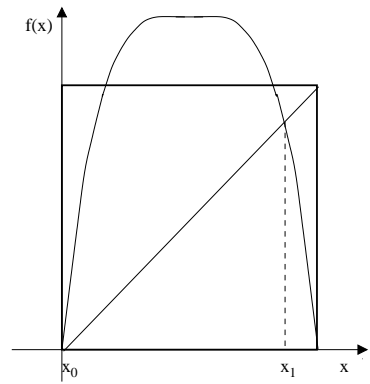,height=5cm}}
\caption{ The map on the [0,1] interval}
\label{figmap}
\end{figure}
The integral in ($\ref{trace}$) containing $-i\epsilon$ in the denominator
has a pole at $x_0$ shifted down from the real axis into the negative
imaginary half plane and
a pole at $x_1$  shifted up into the positive imaginary half plane. At the other integral in
 ($\ref{reint}$) the poles are moved in the opposite direction.
We shall now evaluate the integrals by contour integration.
The chosen contour is shown on figure ($\ref{contour}$).
\begin{figure}[hbt]
\centerline{\strut\psfig{figure=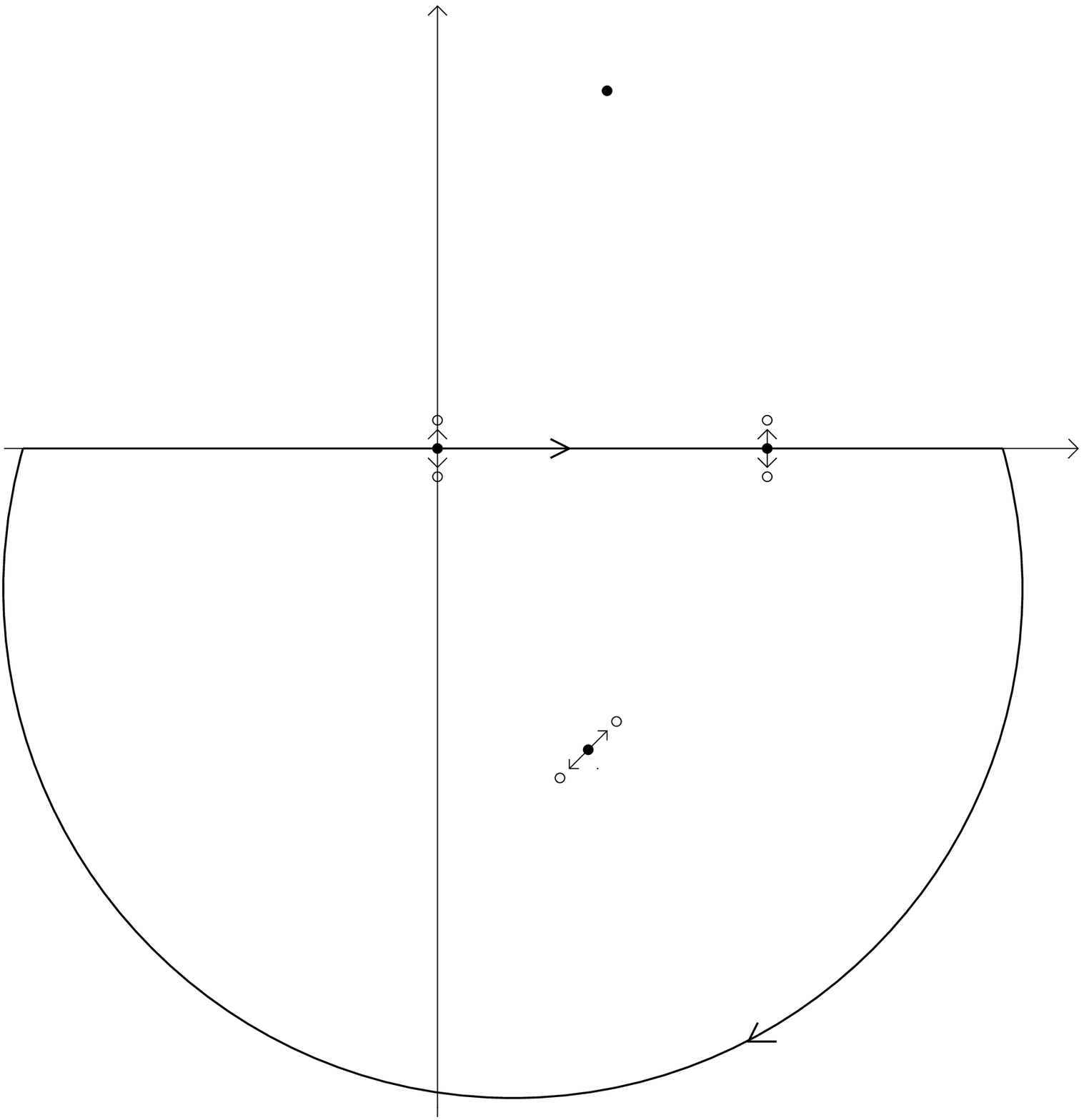,height=6cm}}
\caption{ The chosen contour and the motion of poles. }
\label{contour}
\end{figure}
The integral containing the $-i\epsilon$  will pick up contribution from the fixpoint
 at $x=0$, the other integral from the fixpoint at $x_1$. Both
 will pick up a contribution from the complex fixpoint, but
 they cancel out in the $\epsilon\rightarrow 0$ limit.
 The result of the integration is 
\bea
\mbox{Tr}{\cal L}=-\sum_{n=0}^{\infty}\f{(2n)!\sigma^{2n}}{
2^n n!}\left[\mbox{Res}F_n(x_0)-\mbox{Res}F_n(x_1)
\right],
\label{result1}
\eea
where
\bea
F_n(x)=\f{1}{(x-f(x))^{2n+1}}.
\eea

We used this method to calculate further corrections (around $\sigma^{70}$) to the trace for 
the orbit 0 and orbit 1. This exact method justified the previous approximate calculations to
all known digits.  It turns out that the corrections for orbit 0 
all were positive whereas those for orbit 1 involved sign changes with an apparent period. Furthermore
both were asymptotic series.

\section{Higher order noise corrections}
Practical reasons restricted the calculations with the residue method to around order seventy in the expansion parameter. 
Nevertheless it is possible to give approximate statements for even higher order.

First we transform the residue integrals in ($\ref{reint}$) as
\bea
I_n&=&\f{(2n)!a_{2n}\sigma^{2n}}{2\pi i}\oint dx\f{1}{(x-f(x)\mp i\epsilon)^{2n+1}}=\no \\
& &\f{(2n)!a_{2n}\sigma^{2n}}{2\pi i}\oint
 dx e^{\left(
-(2n+1)\log(x-f(x)\mp i\epsilon)\right)}.
\label{inti}
\eea
 
Using $n$ as a large parameter the integral is next calculated by  the method of steepest descent. The initial loops, $ {\cal C}_0 $
and ${\cal C}_1$, for each fixpoint are expanded until the paths of steepest descents are reached. We refer to \reffig{path}.

\begin{figure}[hbt]
\centerline{\strut\psfig{figure=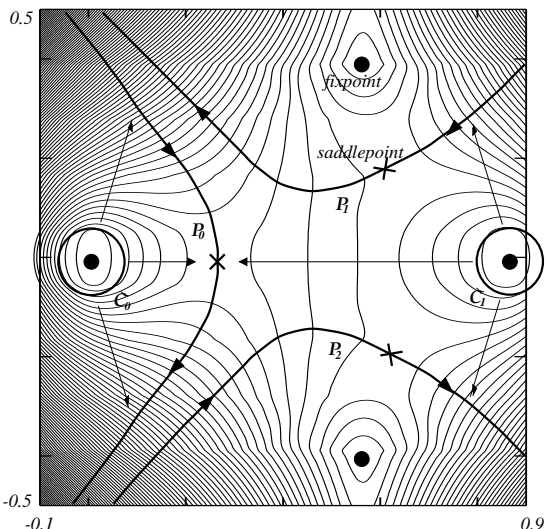,height=7cm}}
\caption{Contourlines of $|x-f(x)| $. Filled circles indicate fixpoints of the map, crosses correspond to saddles and thick lines show paths of steepest descent. }
\label{path}
\end{figure}

The paths of steepest descent pass through the saddle points satisfying
\bea
& &\f{d}{dx}\log(x_s-f(x_s)\mp i\epsilon)=\f{1-f'(x_s)}{x_s-f(x_s)\mp i\epsilon}=0, \no \\
& &f'(x_s)=1.
\eea

Around each saddle we calculate the leading contribution and its corrections as usual.
In the following we will divide the discussion into orbit 0 and orbit 1. 
For our map there are three saddle points: one real and two complex conjugate.

In case of orbit $0$, the integral is carried out on the infinite path ${\cal P}_0$ shown
 on figure \reffig{path}. Here only the saddle point on the real axis contributes.

The result of the integration to leading order is 
\bea
I_n^0&=&\f{(2n)!\sigma^{2n}}{2^{n+1}n!\pi }e^{-(2n+1)\log(x_s-f(x_s))}\sqrt{\f{\pi(x_s-f(x_s))}{
(2n+1)f''(x_s)}}= \no \\
& &\f{2^{n}\Gamma(n+1/2)}{2\pi(x_s-f(x_s))^{2n}\sqrt{2(n+1/2)
(x_s-f(x_s))f''(x_s)}}.\nonumber \\
\label{sadi}
\eea
For a convenient representation we introduce the factors
\bea
C_0&:=&\f{1}{2\pi\sqrt{2(x_s-f(x_s)) f''(x_s))}}, \label{C0} \\
b_0&:=&\f{2}{(x-f(x))^2} \label{b0}
\eea
and the result takes the form  
\bea
I_n^0=C_0\f{\sigma^{2n}b_0^n\Gamma(n+1/2)}{\sqrt{n+1/2}}.
\eea

We show the ratio of consecutive terms $I_{n+1}/I_n$ on \reffig{ratioZero} with the ratios from the exact residue calculation and conclude that the real saddle indeed controls the leading behaviour.

\begin{figure}[hbt]
\centerline{\strut\psfig{figure=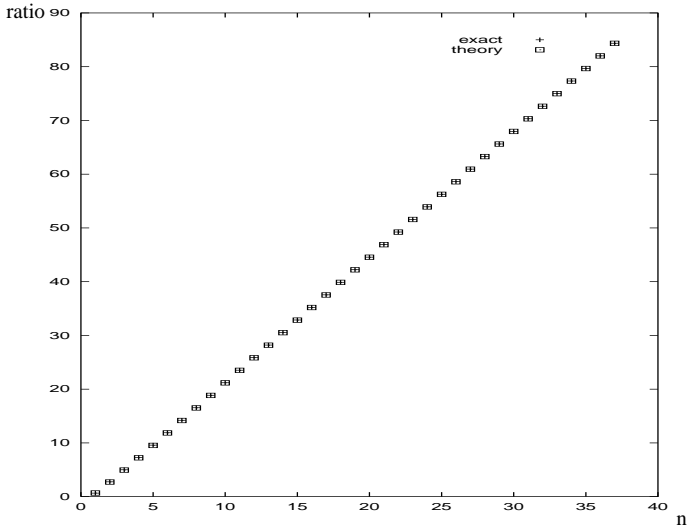,height=7cm}}
\caption{Ratio of noise corrections $\f{I_{2 n + 2}}{I_{2 n}}$ for orbit 0. }
\label{ratioZero}
\end{figure}

For the orbit 1 all three saddles are relevant, since the initial loop ${\cal C}_1$ decomposes into ${\cal P}_0, {\cal P}_1$ and
 ${\cal P}_2$. However, the real saddle now becomes subdominant as can be seen from e.g. the
contour plot \reffig{path}. By analyticity of the map the two complex saddles give results that are complex conjugate to each other.

At the  complex saddle points $(z_1,z_2)$ we introduce similar factors $C_1, b_1$ as in ($\ref{C0},\ref{b0}$) now complex.

The contribution from these two complex saddles to  $I_n^1$ can now be written as
\bea
& & \f{\Gamma(n+1/2)}{\sqrt{n+1/2}}\left[C_1b_1^n+C_1^+(b_1^+)^n\right]= \nonumber \\
& &\f{\Gamma(n+1/2)}{\sqrt{n+1/2}}\left[rq^n\cos(n\phi-\alpha)
\right].
\eea
Here $r,q,\phi$ and $\alpha$ are easily found from  $C_1,b_1$ above.

The final result for orbit 1 also includes the real saddle and  to leading order is given by
\bea
I_n^1=\f{\Gamma(n+1/2)}{\sqrt{n+1/2}}\left[C_0b_0^n+rq^n\cos(n \phi - \alpha)
\right].
\eea

We remark that the same calculation could be done for another distribution than Gaussian. One would find  an identical  functional form  up to a multiplicative constant, the corresponding moment.

\begin{figure}[hbt]
\centerline{\strut\psfig{figure=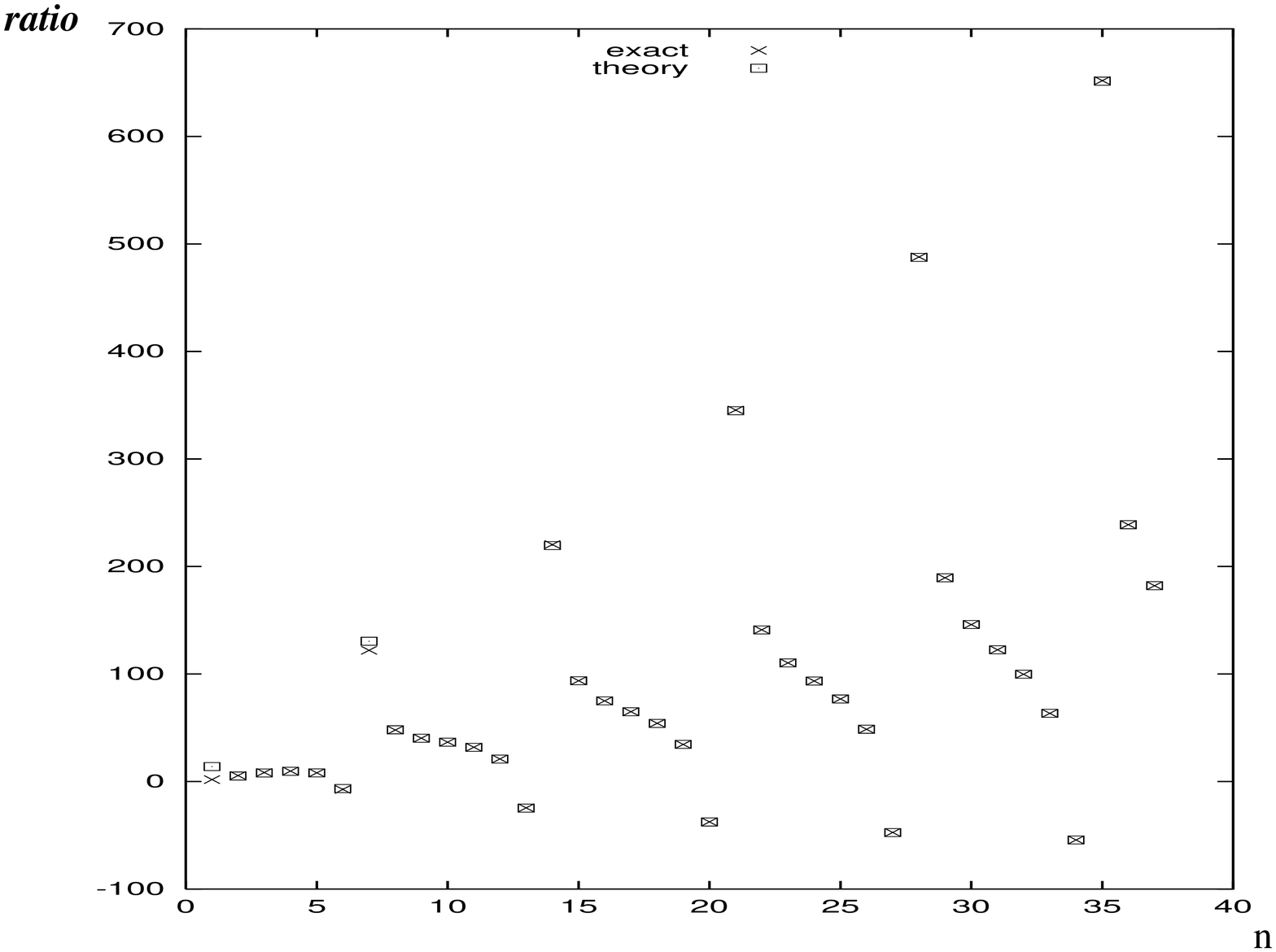,height=7cm}}
\caption{Ratio of noise corrections $\f{I_{2 n + 2}}{I_{2 n}}$ for orbit 1 }
\label{ratioOne}
\end{figure}

Here the plot of the ratios of the consecutive terms \reffig{ratioOne} show a more complex behaviour than for orbit 0. This we attribute to the presence of dominant complex saddles, which will introduce phases and hence the  sign changes already mentioned. The apparent  periodicity of seven is nicely captured by the two complex saddles. Nevertheless, high order calculations actually find deviations from this pattern in agreement with that the phase calculated does not evaluate exactly to an integer multiple of $2 \pi /7.$  The saddle point expansion in general improves  for high orders as can be seen on \reffig{relerror}. For our calculations  two corrections to the complex saddles give the best results in the regime of $n$ we are considering. This is because as an  asymptotic series  truncation is required at an optimal point when $n $ is finite. So adding more corrections will only improve the results for higher $n$ but actually make the lower less accurate.

\begin{figure}[hbt]
\centerline{\strut\psfig{figure=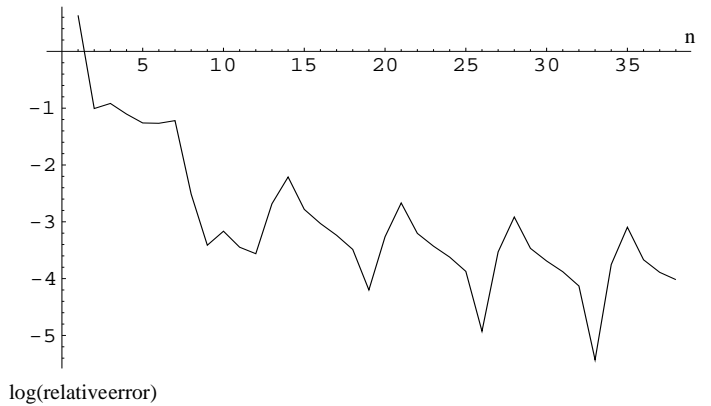,height=7cm}}
\caption{The logarithm of relative error for the  noise corrections for orbit 1 }
\label{relerror}
\end{figure}

\section{Summary and Conclusion }

We have  studied the evolution operator for a discrete Langevin equation with a strongly hyperbolic 
classical dynamics and noise with finite moments. Using a perturbative expansion of the evolution operator we have calculated high order corrections to its trace in the case of a quartic map and Gaussian noise. The leading contributions  come from the period one orbits of the map. We have found the asymptotic behaviour of the corrections using the method of steepest descent. The functional form of the high order corrections have the same form independent of the actual noise distribution up to a constant given by the corresponding moment. The asymptotics of  the trace of the evolution operator are governed by  subdominant terms corresponding to terms previously neglected in the perturbative expansion.

\section{Acknowledgements}

G.V. and G.P. gratefully acknowledges the financial support of
the Hungarian Ministry of Education, FKFP 0159/1999, OMFB, OTKA T25866/F17166.
G.V and A.V were also partially supported by the French Minist\`ere des
 Affaires \'Etrang\`eres.
N.S. is supported by the Danish Research Academy Ph.D.
fellowship.

%REFERENCES -------------------------------------------------------


\begin{thebibliography}{999}

\bibitem{noisy_Fred}  P. Cvitanovi\'c, C.P.~Dettmann, R.~Mainieri and G. Vattay,
        % {\em Trace formulas for stochastic evolution operators:
        % Weak noise perturbation theory},
        {\em J. Stat. Phys. \bf 93}, 981 (1998);
        {\tt chao-dyn/9807034}.

\bibitem{conjug_Fred} P. Cvitanovi\'c, C.P.~Dettmann, R.~Mainieri and G. Vattay,
        {\em Trace formulas for stochastic evolution operators:
         Smooth conjugation method},
        submitted to {\em Nonlinearity} (November 1998);
        {\tt chao-dyn/9811003}.

\bibitem{diag_Fred}  P. Cvitanovi\'c, C.P.~Dettmann, N.Sondergaard,G. Vattay and G. Palla
        {\em Spectrum of Stochastic Evolution Operators: Local Matrix Representation Approach}
	{\em Phys. Rev. E {\bf 60}, 3936-3941 (October 1999)}

\bibitem{bene} J. Bene and P. Sz\'epfalusy, Phys. Rev. A {\bf 37}, 871 (1988).


\bibitem{gutbook} M.C. Gutzwiller, {\em Chaos in Classical
	and Quantum Mechanics} (Springer, New York 1990).

%\bibitem{abramo} M. Abramowitz and I. A. Stegun,
%          {\em Handbook of mathematical functions with formulas, graphs
%and mathematical tables}, Chapter 24, page 823,
%formula I.B, Dover, New York (1972).



\bibitem{QCcourse} P. Cvitanovi\'c, et al.,
	{\em Classical and Quantum Chaos},
	{\tt http://www.nbi.dk/ChaosBook/}, 
	(Niels Bohr Institute, Copenhagen 1999).

%\bibitem{Rugh92} H.H. Rugh,
%        ``The Correlation Spectrum for Hyperbolic Analytic Maps'',
%	{\em Nonlinearity \bf 5}, 1237 (1992).

%\bibitem{VatBS}
%	 G. Vattay and P.E. Rosenqvist,
%       %{\em ``Periodic Orbit Quantization beyond Semiclassical Approximation"},
%        {\em Phys. Rev. Lett. \bf 76}, 335 (1996),
%        {\tt chao-dyn/9509015};
%	G. Vattay,
%        {\em ``Bohr Sommerfeld Quantization of Periodic Orbits"},
%        {\em Phys. Rev. Lett. \bf 76}, 1059 (1996).

\bibitem{Dingle} R.B. Dingle,
        {\em Asymptotic Expansions: their Derivation and Interpretation}
        (Academic Press, London, 1973).

%\bibitem{asym_Fred} P. Cvitanovi\'c, C.P.~Dettmann, G. Palla,
%	N. S\o ndergaard and G. Vattay,
%	{\em Trace formulas for stochastic evolution operators:
%	 Beyond all orders},
%	in preparation.


%%%%%%%%%%NOT USED:


\end{thebibliography}
\end{document}